\documentclass[doublecol]{epl2}

\usepackage{graphicx}

\title{A dissipative force between  colliding viscoelastic bodies: \\
 Rigorous approach}
\shorttitle{A dissipative force between viscoelastic bodies}

\author{Nikolay V.\ Brilliantov~\inst{1} \and Anastasiya V.\ Pimenova~\inst{2} \and Denis S.\ Goldobin~\inst{1,2,3}}
\shortauthor{N.\ Brilliantov \etal}

\institute{
  \inst{1} Department of Mathematics, University of Leicester, Leicester LE1 7RH, United Kingdom \\
  \inst{2} Institute of Continuous Media Mechanics, UB RAS,
             Perm 614013, Russia \\
  \inst{3} Theoretical Physics Department, Perm State University,
             Perm 614990, Russia
}
 {Collisions }

\abstract{A collision of viscoelastic bodies is analysed within a mathematically rigorous approach. We develop a perturbation scheme to solve continuum mechanics equation, which deals simultaneously with strain and strain rate in the bulk of the bodies' material. We derive dissipative force that acts between particles and express it in terms of particles' deformation, deformation rate and material parameters. It differs noticeably from the currently used dissipative force, found within the quasi-static approximation and does not suffer from inconsistencies of this approximation. The proposed approach may be used for other continuum mechanics problems where the bulk dissipation is addressed. }

\begin{document}
\maketitle
\section{Introduction}

Granular materials are abundant in nature and play an important role in industry. Properties of these systems are very unusual and depend on the applied load: for a small load a granular medium  behaves  as a solid, for a larger load it flows  like a liquid, while at still larger  excitations, a gas-like behavior may be observed. Such rich behavior is a consequence of the dissipative nature of the interaction forces between particles comprising a granular system. Therefore for an adequate description of granular media it is crucial to develop a quantitative model for the dissipative forces at particles' contacts.

While the elastic component of the inter-particle force is known for more than a century from the famous work of Hetrz~\cite{hertz1881}, where a mathematically rigorous theory has been developed, a rigorous derivation for the dissipative component is still lacking. The existing phenomenological expressions for the dissipative force used either linear, e.g.~\cite{PoeschelSchwager2005,LudingReview2008} or quadratic~\cite{Poschl1928} dependence
on the deformation rate; these however do not agree with the experimental data, e.g.~\cite{PoeschelSchwager2005,Poeschel2011}. An attempt to obtain a dissipative  force from the basic  principles, has been undertaken in \cite{Pao1955}; only a limited class of deformations has been addressed there.

A first complete derivation of the dissipative force between viscoelastic bodies from the continuum mechanics equations  has been done only recently~\cite{bshp96}. In this work a so called \emph{quasi-static} approximation has been introduced. The functional dependence of the dissipative force on the deformation and deformation rate, found in Ref.~\cite{bshp96}, has been already proposed (without any mathematical derivation) in the earlier work of Kuwabara and Kono~\cite{kk87}. In later studies~\cite{Zheng1,Zheng2} a flaw in the  derivation of the dissipative force in  Ref.~\cite{bshp96} was corrected; still the restrictive assumption of the quasi-static approximation was used~\cite{Zheng1,Zheng2}.

In the quasi-static approximation it is assumed that the
displacement field in the deformed material completely  coincides
with that for the static case. That is,  an immediate response of
the particles' material to the external load is supposed. More
precisely, the  quasi-static approximation implies that: (i) the
characteristic deformation rate is much smaller than the speed of
sound in the system and (ii) the microscopic relaxation time of
the particle's material is negligibly small as compared to the
duration of the impact. The precise definition of the former
quantity will be given below, physically, however  it
characterizes the response of the material to the applied load. In
the present study we develop a \emph{mathematically rigorous}
perturbative approach, which allows to go beyond the quasi-static
approximation; we demonstrate that this approximation, although
being physically plausible, is not mathematically complete. This
happens because the deviations  from the static deformations,
neglected in the quasi-static approximation, ultimately yield a
contribution to the dissipative force, comparable to the force
itself in this approximation. The proposed approach may be also
used to analyze other time-dependent impact problems.

\section{Perturbation scheme for the continuum mechanics equation}
\label{sec:Int_forces}
To find a force acting between the bodies in a contact with a
given deformation at their surfaces, one needs to solve continuum
mechanics equation for the stress tensor. Integration of the
obtained stress over the contact area yields  the inter-particle
force. The contact mechanics equation, that is, the equation of
motion for a body material, generally reads,
e.g.~\cite{Landau:1965},
\begin{equation}
\label{eq:1}
\rho \ddot{\bf u} = {\bf \nabla } \cdot \hat{\sigma}= {\bf \nabla } \cdot  \left( \hat{\sigma}^{ el} + \hat{\sigma}^{ v} \right) \, ,
\end{equation}
where $\rho$ is the material density,  ${\bf u}={\bf u}({\bf r})$ is the displacement field in a point ${\bf r}$ and $\hat{\sigma}$ is the stress tensor, comprised of the elastic $\hat{\sigma}^{ el}$ and viscous $\hat{\sigma}^{ v}$ parts. The elastic stress linearly depends on the strain tensor $u_{ij}=\frac12 \left(\nabla_i u_j+\nabla_j u_i \right) $~\cite{Landau:1965},
\begin{equation}
 \label{eq:sigma__el}
 \sigma^{el}_{ij} ({\bf u})=2E_1
\left(u_{ij} -\frac13 \delta_{ij} u_{ll}\right) +E_2 \delta_{ij}
u_{ll}\, ;
\end{equation}
correspondingly,  the viscous stress depends on the strain \emph{rate} tensor~\cite{Landau:1965}:
\begin{equation}
 \label{eq:sigma_vis}
 \sigma^{v}_{ij} (\dot{\bf u})=2\eta_1
\left(\dot{u}_{ij} -\frac13 \delta_{ij} \dot{u}_{ll}\right) +\eta_2 \delta_{ij}
\dot{u}_{ll}\, .
\end{equation}
Here $E_1= \frac{Y}{2(1+\nu)}$, $E_2= \frac{Y}{3(1-2\nu)}$, with $Y$ and $\nu$ being respectively the Young modulus and Poisson ratio, and $\eta_1$ and $\eta_2$ are respectively shear and bulk viscosities  of the bodies' material; $i,j,l$ denote Cartesian coordinates and the Einstein's summation rule is applied.

Let us estimate the magnitude of the different terms in Eq.(\ref{eq:1}). This may be easily done using the dimensionless units. For the length scale we take $R$, which corresponds to the characteristic size of colliding bodies, while for the time scale we use the collision duration $\tau_c$.  Then $v_0=R/\tau_c$ is the characteristic velocity at the impact. Taking into account that differentiation with respect to a coordinate yields for dimensionless quantities the factor $1/R$ and with respect to time the factor $1/\tau_c$, we obtain

\begin{eqnarray}
\label{eq:2a}
\nabla \sigma^{v} & \sim &  \lambda_1 \,\nabla \sigma^{el} \qquad \qquad \lambda_1 = \tau_{rel}/ \tau_c \\
\label{eq:2b}
\rho \ddot{ u}  & \sim &  \lambda_2 \, \nabla \sigma^{el} \qquad \qquad \lambda_2 = v_0^2/c^2\,.
\end{eqnarray}
Here $c^2=Y/\rho$ and $\tau_{rel}=\eta/Y$ characterize
respectively the speed of sound and the microscopic relaxation
time in the material and $\eta \sim \eta_{1}\sim
\eta_{2}$~\cite{bshp96}. Hence, the term associated with
the viscous stress is smaller by factor $\lambda_1$  than the one
corresponding to  the elastic stress, while the term associated
with the inertial effects is smaller by the  factor $\lambda_2$.

\begin{figure}[t]
\center{
 \includegraphics[width=0.3\textwidth]%
 {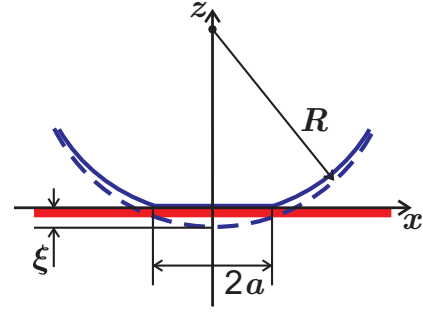}}

  \caption{
Collision of visco-elastic sphere of radius $R$ with undeformable plane. Here $a$ is the radius of the contact zone and $\xi$ is the  deformation. The contact plane is located at $z=0$.
 }
  \label{fig1}
\end{figure}

Neglecting terms, of the order of $\lambda_1$ and $\lambda_2$,
that is, the terms $\nabla \sigma^{v}$ and $\rho \ddot{ u} $,
Eq.~(\ref{eq:1}) simplifies to
\begin{equation}
\label{eq:2c}
{\bf \nabla} \cdot \hat{\sigma}^{el} ({\bf u})=0,
\end{equation}
which yields the static displacement field ${\bf u} ={\bf u}({\bf r})$. This approximation corresponds to the quasi-static approximation, used in the literature \cite{bshp96,bri2,Dintwa2004,Zheng1,Zheng2}. Neglecting terms of the order $\lambda_2$ but keeping terms of the order of $\lambda_1$ yields:
\begin{equation}
\label{eq:3}
{\bf \nabla}  \cdot \hat{\sigma}= {\bf \nabla}  \cdot \left( \hat{\sigma}^{el} ({\bf u})+ \hat{\sigma}^{v} (\dot{{\bf u}}) \right)=0.
\end{equation}
\noindent Physically, the above equation describes the over-damped
motion of a material when the inertial effects,  proportional to
$\lambda_2$, are negligible; the excitation of elastic waves in
this case may be ignored. Such conditions are important for many
applications, especially for slow collisions.

To go beyond the quasi-static approximation one  has to solve Eq.~(\ref{eq:3}) which contains both the displacement field ${\bf u}$ as well as its time derivative, $\dot{{\bf u}}$. Eq.~(\ref{eq:3}) needs to be supplemented by the boundary conditions. These correspond to vanishing stress on the free surface of the bodies and given displacement ${\bf u}$ at  the contact area. For simplicity we consider here a collision of a sphere of radius $R$ with a hard undeformable plane located at $z=0$, Fig.~\ref{fig1}. The generalization for a  contact of two arbitrary convex bodies of different materials is straightforward, but leads to cumbersome notations; it will be addressed elsewhere \cite{Denisetal2014}. Let  $\xi=R-z_O$ be the deformation, where $z_O$ is $z$-coordinate of the center of mass of the sphere, then $z$-component of the displacement on the contact plane reads for small deformations~\cite{Landau:1965}:
\begin{equation}
\label{eq:3a}
u_z(x,y)=\xi -\frac{1}{2R}(x^2+y^2)\,.
\end{equation}
In a vast majority of applications $\lambda_1= \tau_{rel}/\tau_c
\ll 1$, which implies that the viscous stress is small as compared
to the elastic stress. This allows to solve Eq.~(\ref{eq:3})
perturbatively, as a series in a small parameter $\lambda_1
\propto \eta$. Here we follow the standard perturbation scheme,
e.g.~\cite{book}: To notify the order of different terms,  we
introduce a ``technical'' small parameter $\lambda$, which at the
end of computations is to be taken as unity.  Hence one can write,
\begin{equation}
\label{eq:4a}
{\bf u}({\bf r})= {\bf u}^{(0)}({\bf r})+\lambda {\bf u}^{(1)}({\bf r})+ \lambda^2 {\bf u}^{(2)}({\bf r}) +\ldots
\end{equation}
and
\begin{equation}
\label{eq:4b}
\hat{\sigma} = \hat{\sigma}^{el} + \lambda \hat{\sigma}^{v}
\end{equation}
Substituting the Eqs.~(\ref{eq:4a}) and (\ref{eq:4b}) into
Eq.~(\ref{eq:3}) and collecting terms of the same order in
$\lambda$ yields a hierarchic set of equations.  The zero-order
equation reads,
\begin{eqnarray}
\label{eq:5}
&&{\bf \nabla}  \cdot  \hat{\sigma}^{el} \left({\bf u}^{(0)}\right)=0 \\
&&\left. u_z^{(0)}\right|_{z=0}=\xi -\frac{1}{2R}(x^2+y^2)\,.
\nonumber
\end{eqnarray}
The first-order (that is, proportional to $\lambda$) equation is
\begin{eqnarray}
\label{eq:6}
&&{\bf \nabla}  \cdot \left( \hat{\sigma}^{el} ({\bf u}^{(1)})+ \hat{\sigma}^{v} (\dot{{\bf u}}^{(0)}) \right)=0 \\
&& \left. u_z^{(1)}\right|_{z=0}=0 \, , \nonumber
\end{eqnarray}
and so on, where the expressions for $\hat{\sigma}^{el}$ and $\hat{\sigma}^{v}$ are given by Eqs.~(\ref{eq:sigma__el}) and (\ref{eq:sigma_vis}). In all these equations the stress tensor vanishes on the free surface. Note that in the proposed perturbation scheme, only zero-order equation (\ref{eq:5}) has non-zero boundary conditions, corresponding to the boundary conditions (\ref{eq:3a}) of the initial problem; all other, high-order  perturbation equations, have homogeneous boundary conditions. Such partition of the boundary conditions is justified due to linearity of the problem.

The zero-order solution ${\bf u}^{(0)}$ of Eq.~(\ref{eq:5}) of the
above perturbative approach is to be substituted into
Eq.~(\ref{eq:6}) to find the first-order solution ${\bf u}^{(1)}$,
which may be further used to obtain ${\bf u}^{(2)}$ from the
second-order equation, etc. Hence the series
\begin{equation}
\label{eq:4}
\hat{\sigma} = \hat{\sigma}^{(0)} + \lambda \hat{\sigma}^{(1)} + \lambda^2 \hat{\sigma}^{(2)} + \ldots
\end{equation}
is generated, where $\hat{\sigma}^{(0)}=\hat{\sigma}^{el\,(0)}
=\hat{\sigma}^{el} \left({\bf u}^{(0)}\right)$ is the  zero-order
term, $\hat{\sigma}^{el\,(1)}= \hat{\sigma}^{el} \left({\bf
u}^{(1)}\right)$ and $\hat{\sigma}^{v\,(1)} =\hat{\sigma}^{v}
\left(\dot{\bf u}^{(0)}\right)$ are the first-order terms,
$\hat{\sigma}^{el\,(2)} =\hat{\sigma}^{el} \left({\bf
u}^{(2)}\right)$ is the second-order term with respect to
$\lambda_1 \propto \eta$, etc.

\section{Zero-order solution. Hertz theory}

To illustrate the approach we start with the zero-order Eq.~(\ref{eq:5}). It corresponds to the quasi-static approximation (\ref{eq:2c}), which solution is known. In the above notations Eq.~(\ref{eq:5}) reeds
\begin{equation}
\nabla_j\sigma_{ij}^{el\,(0)}
  =E_1\Delta u_i^{(0)}+\left(E_2+\frac{1}{3}E_1\right)\nabla_i \nabla_j u_j^{(0)}=0\,.
\label{eq04}
\end{equation}
To solve Eq.~(\ref{eq04}) we use the approach of Ref.~\cite{Landau:1965} and write the  solution as
\begin{equation}
{\bf u}^{(0)}=f^{(0)}{\bf e}_z+\nabla\varphi^{(0)}\,,
\label{eq05}
\end{equation}
where  $\varphi^{(0)}=K^{(0)}zf^{(0)}+\psi^{(0)}$,    $K^{(0)}$ is
some constant to be found and  $f^{(0)}$ and  $\psi^{(0)}$  are
unknown harmonic functions ($\Delta f^{(0)}=0$ and $\Delta
\psi^{(0)}=0$). We assume the lack of tangential stress at the
interface, which is e.g. fulfilled when the bodies at a contact
are of the same material. Physically, the substitute (\ref{eq05})
is dictated by the symmetry of the problem: The main displacement
of the material occurs along $z$-axes. Taking into account that
\begin{equation}
\displaystyle\Delta {\bf u}^{(0)}=\Delta\nabla\varphi^{(0)}
 =2K^{(0)}\nabla\frac{\partial f^{(0)}}{\partial z}
\label{eq:5a}
\end{equation}
and
\begin{equation}
 \displaystyle\nabla\cdot {\bf u}^{(0)}=(1+2K^{(0)})\frac{\partial f^{(0)}}{\partial z},
\label{eq:5b}
\end{equation}
as it follows from Eq.~(\ref{eq05}), we recast Eq.~(\ref{eq04})
into the form:
\begin{eqnarray}
\nabla_j\sigma_{ij}^{el(0)}\!\! &=& \!\! \left[2E_1 K^{(0)} + \right. \\
   &+&\left. \!\!(1+2K^{(0)})\left(E_2+\frac{E_1}{3}\right)\right]
 \nabla_i\frac{\partial f^{(0)}}{\partial z}=0\,,\nonumber
\end{eqnarray}
which implies (for non-zero $f^{(0)}$) that
\begin{equation}
K^{(0)}=-\frac{1}{2}\frac{3E_2+E_1}{3E_2+4E_1}\,.
\label{eq07}
\end{equation}
Consider now the boundary condition for the stress tensor.
Obviously, on the free boundary all components of the stress
vanish. In the contact region, located at the surface, $z=0$, the
tangential components of the stress tensor $\sigma_{zx}$ and
$\sigma_{zy}$ vanish as well, while the normal component of the
stress tensor equals (up to the sign) to the normal component of
the external pressure $P_z^{(0)}$, e.g.~\cite{Landau:1965}:
\begin{equation}
\left.\sigma_{zx}^{el(0)}\right|_{z=0}\!=\!0; \quad \left.\sigma_{zy}^{el(0)}\right|_{z=0}\!=\!0; \quad \left.\sigma_{zz}^{el(0)}\right|_{z=0}\!=\!-P_z^{(0)}.
\label{eq10}
\end{equation}
Using the Eq.~(\ref{eq:sigma__el}) for the elastic part of the stress tensor, together with the displacement vector (\ref{eq05}) we recast the boundary conditions (\ref{eq10}) into the form:
\begin{eqnarray}
\label{eq:12}
&&\left.\frac{\partial}{\partial x}\!\left(\frac{3E_1 }{4E_1+3E_2}f^{(0)}
 \!+\!2\frac{\partial\psi}{\partial z}\right)\right|_{z=0} \!\!=0 \\
\label{eq:13}
&&\left.\frac{\partial}{\partial y}\!\left(\frac{3E_1 }{4E_1+3E_2}f^{(0)}
 \!+\!2\frac{\partial\psi}{\partial z}\right)\right|_{z=0} \!\!=0 \\
\label{eq14}
&&\left.\frac{\partial}{\partial z}\!\left(\frac{3E_1 }{4E_1 +3E_2}f^{(0)}
 \!+\!2\frac{\partial\psi}{\partial z}\right)\right|_{z=0}\!\!
 \!=\!- \frac{P_z^{(0)}}{E_1}.
\end{eqnarray}
From Eqs.~(\ref{eq:12}) and (\ref{eq:13}) follows the relation between $f^{(0)}$ and
 $\frac{\partial\psi}{\partial z}$ at $z=0$:
\begin{equation}
\label{eq14a}
\left.\left(\frac{\partial\psi}{\partial z}
 +\frac{3}{2}\frac{E_1}{4E_1+3E_2}f^{(0)}\right)\right|_{z=0}
 ={\rm const} =0\,.
\end{equation}
The constant in the above relation equals to zero, since it holds
true independently on the coordinate that is, also at the
infinity; at the infinity, however, the deformation and thus the
above functions vanish. Since $f^{(0)}$, $\psi$ and
$\partial\psi/\partial z$ are harmonic functions, the
condition that their  linear combination  vanishes on the
boundary, Eq.~(\ref{eq14a}), implies that this combination is zero in the total
domain, that is,
\begin{equation}
\frac{\partial\psi}{\partial z}
 =-\frac{3}{2}\frac{E_1}{4E_1+3E_2}f^{(0)}\,.
\label{eq15}
\end{equation}
Substituting the last relation into (\ref{eq14}) yields
\begin{equation}
\left.\frac{\partial f^{(0)}}{\partial z}\right|_{z=0}
 =-\frac{4E_1+3E_2}{E_1(E_1 +3E_2)}P_z^{(0)}   .
\label{eq16}
\end{equation}
Since $f^{(0)}$ is a harmonic function, one can use the relation between the normal derivative of a harmonic function on a  surface and its value in the bulk, as it follows from the theory of harmonic functions (see e.g.~\cite{Tikhonov,Landau:1965}), hence we find:

\begin{equation}
f^{(0)}({\bf r})=\frac{4E_1+3E_2}{2\pi E_1(E_1+3E_2)}
 \int\!\!\!\int_S\frac{ P_z^{(0)}(x',y')\,\mathrm{d}x'\mathrm{d}y'}{ \left|{\bf r}
 -{\bf r}'\right|}\,,
\label{eq17}
\end{equation}
where $S$ is the contact area.

Using Eq.~(\ref{eq05})  we can write $z$-component of the zero-order displacement at $z=0$ as
$$
\displaystyle\left.u_z^{(0)}\right|_{z=0}=(1+K^{(0)})\left.f^{(0)}\right|_{z=0}
 +\left.\frac{\partial\psi}{\partial z}\right|_{z=0},
$$
which together with (\ref{eq15}) and definition of  $K^{(0)}$
(Eq.~(\ref{eq07})) yields,
\begin{equation}
\left.u_z^{(0)}\right|_{z=0}=\frac{1}{2}\left.f^{(0)}\right|_{z=0} \,.
\label{eq20}
\end{equation}
If we now express $E_1$ and $E_2$  in terms of $\nu$ and $Y$, we obtain from Eqs.~(\ref{eq20}),  (\ref{eq17}) and (\ref{eq10}):
\begin{equation}
\label{eq20a}
\left.u_z^{(0)}\right|_{z=0}=-\frac{(1-\nu^2)}{\pi Y} \int\!\!\! \int_S
\frac{ \sigma_{zz}^{el(0)}(x',y', z=0)\,\mathrm{d}x'\mathrm{d}y'}{ \left|{\bf r} -{\bf r}'\right|}\,.
\end{equation}
Eq.~(\ref{eq20a}) is a  standard relation of the  static continuum
theory, e.g.~\cite{Landau:1965}. Physically it relates the
distribution of the normal displacement and normal stress at  the
contact zone. The distribution of the normal pressure there
follows from the Hertz theory (see e.g.~\cite{Landau:1965}):
\begin{equation}
\left.-\sigma_{zz}^{el(0)}\right|_{z=0} = P_z^{(0)}
 =\frac{2Y}{\pi R (1-\nu^2)} \sqrt{a^2 -(x^2+y^2)}\,,
\label{eq18}
\end{equation}
where $a$ is the radius of the contact circle. Substituting
Eq.~(\ref{eq18}) into (\ref{eq20a}) and performing integration
over the contact zone we obtain, as expected, the displacement
(\ref{eq:3a}). Moreover, since $\xi=\left.
u_z^{(0)}(x=0,y=0)\right|_{z=0}$, we find  the relation between
deformation and the radius of the contact circle, $\xi =a^2/R$.
Integrating the stress  (\ref{eq18}) over the contact we obtain
the elastic Hertzian force, e.g.~\cite{Landau:1965}:

\begin{equation}
F_H= F_z^{el(0)}= B\xi^{3/2},  \qquad \quad B=\frac{4Y \sqrt{R}}{3(1-\nu^2)}.
\label{eq:FH}
\end{equation}
Obviously, the zero-order terms refer to the static case and do
not describe dissipation. As it follows from the above discussion
[see Eq.~(\ref{eq:6})] there are two first-order terms,
$\hat{\sigma}^{v} \left(\dot{\bf u}^{(0)}\right) =
\hat{\sigma}^{v\,(1)}$ and $\hat{\sigma}^{el} \left({\bf
u}^{(1)}\right) = \hat{\sigma}^{el\,(1)}$. The former one depends
on the  \emph{known} zero-order solutions ${\bf u}^{(0)}({\bf r })
$ and hence may be related to the  zero-order stress
$\hat{\sigma}^{el(0)}= \hat{\sigma}^{el}({\bf u}^{(0)})$ as
\begin{equation}
\sigma_{ij}^{v(1)}=\frac{\eta_1}{E_1}\dot{\sigma}_{ij}^{el(0)}
 +\left(\eta_2-\eta_1\frac{E_2}{E_1}\right)(1+2K^{(0)})
 \frac{\partial\dot{f}^{(0)}}{\partial z}\delta_{ij}\,,
\label{eq09}
\end{equation}
where we use Eqs.~(\ref{eq:sigma__el}), (\ref{eq:sigma_vis}) and (\ref{eq:5b}). If we now apply Eq.~(\ref{eq16}) for $\partial{f}^{(0)}/\partial z$ and Eq.~(\ref{eq07}) for the constant $K^{(0)}$, we find the $zz$-component of this tensor at the contact plane, $z=0$:
\begin{eqnarray}
\label{eq19}
\sigma_{zz}^{v(1)}(x,y,0)\!\!\!&=& \!\! \!\alpha_0  \dot{\sigma}_{zz}^{el(0)}(x,y,0) \\
\label{eq19a}
\alpha_0 \!\!\! &=& \!\! \! \frac{3\eta_2+\eta_1}{3E_2 +E_1} \! = \! \frac{(2+2\nu)(1-2\nu)(3\eta_2+\eta_1)}{3Y} \nonumber
\end{eqnarray}
where the definitions of $E_1$ and $E_2$ have been used.

The other first-order term, $\hat{\sigma}^{el} \left({\bf
u}^{(1)}\right) = \hat{\sigma}^{el\,(1)}$ depends on the
first-order displacement ${\bf u}^{(1)}({\bf r})$ which is still
to be found. Neglecting this term and keeping only one first-order
term (\ref{eq19}) corresponds to the quasi-static approximation
for the dissipative force~\cite{bshp96} discussed above. The
expression for $\alpha_0$ coincides with the result
of~\cite{Zheng1,Zheng2}, where the necessary corrections have been
implemented.

\section{First-order solution. Beyond the quasi-static approximation}

Turn now to the first-order equation (\ref{eq:6}), which is
actually an equation for  the function ${\bf u}^{(1)}$ that
describes deviations of the displacement from the static case. We
write this equation as
\begin{equation}
\nabla_j\sigma_{ij}^{el(1)}=-\nabla_j \sigma_{ij}^{v(1)}\,,
\label{eq21}
\end{equation}
where the left-hand side contains the unknown function ${\bf
u}^{(1)}$, while the right-hand side depends on ${\bf u}^{(0)}$
and is therefore known. Using  Eqs.~(\ref{eq:5a}), (\ref{eq:5b})
and Eq.~(\ref{eq07}) for $K^{(0)}$ we obtain for the r.h.s. of
Eq.~(\ref{eq21}):
\begin{eqnarray}
\nabla _j\sigma_{ij}^{v(1)}&=&\left[2\eta _1 K^{(0)}
 +(1+2K^{(0)})\left(\eta _2+\frac{\eta _1}{3}\right)\right]
 \nabla_i\frac{\partial\dot{f}^{(0)}}{\partial z} \nonumber \\
 &=& \frac{3(E_1\eta_2 -E_2 \eta_1)}{(4 E_1+3E_2)} \nabla_i\frac{\partial\dot{f}^{(0)}}{\partial z}\,.
\label{eq08}
\end{eqnarray}
To proceed with the solution of Eq.~(\ref{eq21}) for ${\bf
u}^{(1)}$ we reduce it to the solution of two simpler equations.
Namely, due to linearity of the problem, one can represent the
first-order displacement field as a sum of two parts, ${\bf
u}^{(1)} = \bar{\bf u}^{(1)}+\tilde{\bf u}^{(1)}$, which
correspond to the two parts of the elastic stress tensor,
$\sigma_{ij}^{el(1)}=\tilde{\sigma}_{ij}^{el(1)}+\bar{\sigma}_{ij}^{el(1)}$.
Here the first part of $\sigma_{ij}^{el(1)}$ is the solution of
the \emph{inhomogeneous} equation with homogeneous boundary
conditions:
\begin{eqnarray}
\label{eq22}
&&\nabla_j\tilde{\sigma}_{ij}^{el(1)}=-\nabla_j\sigma_{ij}^{v(1)}  \\
\label{eq22a}
&&\left.\tilde{\sigma}_{xz}^{el(1)}\right|_{z=0}=
 \left.\tilde{\sigma}_{yz}^{el(1)}\right|_{z=0}=
 \left.\tilde{\sigma}_{zz}^{el(1)}\right|_{z=0}=0,
\end{eqnarray}
while the second part is the solution of the \emph{homogeneous} equation with the given boundary conditions for the displacement $\bar{u}_z^{(1)}$ at the contact plane:
\begin{eqnarray}
\label{eq23}
\nabla_j\bar{\sigma}_{ij}^{el(1)}&=&0 \\ 
\bar{u}_z^{(1)}&=&{u}_z^{(1)}-\tilde{u}_z^{(1)} =-\tilde{u}_z^{(1)} \nonumber \,.
\end{eqnarray}
Here we use the boundary conditions (\ref{eq:6}), that is, $\left. {u}_z^{(1)}\right|_{z=0}=0$. The boundary problem (\ref{eq23}) is exactly the same as the above problem (\ref{eq:5}) for the zero-order functions. Hence the same relation (\ref{eq20a}) holds true for the first-order functions, that is,
\begin{equation}
\label{eq:29b}
\left.\bar{u}_z^{(1)}\right|_{z=0}=-\frac{(1-\nu^2)}{\pi Y} \int\!\!\! \int_S
\frac{ \bar{\sigma}_{zz}^{el(1)}(x',y', z=0)\,\mathrm{d}x'\mathrm{d}y'}{ \left|{\bf r} -{\bf r}'\right|}\,. \nonumber
\end{equation}

To solve Eq.~(\ref{eq22}) we write the displacement field ${\tilde{\bf u}}^{(1)}$ in a form, similar to this for the zero-order solution (\ref{eq05}):
\begin{equation}
{\bf \tilde{u}}^{(1)}=f^{(1)}{\bf e}_z+\nabla\varphi^{(1)}\,,
\label{eq24}
\end{equation}
where $\varphi^{(1)}=K^{(1)}zf^{(1)}+\psi^{(1)}$, $K^{(1)}$ is some constant and  $f^{(1)}$ and  $\psi^{(1)}$ are harmonic functions. Then we can write the first-order elastic stress tensor $\tilde{\sigma}_{ij}^{el(1)}$ as
\begin{eqnarray}
 \tilde{\sigma}_{ij}^{el(1)}&=&(1+2K^{(1)})
 \bigg[E_1(\delta_{jz}\nabla_i f^{(1)}+\delta_{iz}\nabla_jf^{(1)}) \nonumber \\
 &+& \left(E_2-\frac{2}{3}E_1\right)\frac{\partial f^{(1)}}{\partial z}\delta_{ij}\bigg]
 {}+2E_1K^{(1)}z\nabla_i\nabla_j f^{(1)} \nonumber \\
 &+& 2E_1\nabla_i\nabla_j\psi^{(1)}.
\label{eq25}
\end{eqnarray}
Choosing  $K^{(1)}=-\frac12$ the above stress tensor simplifies to

\begin{equation}
\label{eq:25a}
\tilde{\sigma}_{ij}^{el(1)}= - z E_1\nabla_i\nabla_j f^{(1)} + 2 E_1 \nabla_i\nabla_j\psi^{(1)}
\end{equation}
and the boundary conditions (\ref{eq22a}) read:
\begin{eqnarray}
\label{eq:25b}
&& \left.\tilde{\sigma}_{xz}^{el(1)}\right|_{z=0}= \frac{\partial }{\partial x} \left.\left( \frac{\partial \psi^{(1)}}{\partial z} \right) \right|_{z=0}=0 \\
&&\left. \tilde{\sigma}_{yz}^{e(1)}\right|_{z=0}= \frac{\partial }{\partial y} \left.\left( \frac{\partial \psi^{(1)}}{\partial z} \right) \right|_{z=0}=0 \,.
\end{eqnarray}
Therefore we conclude,
\begin{equation}
\label{eq:25c}
\left.\frac{\partial \psi^{(1)}}{\partial z}  \right|_{z=0}={\rm const}=0 \,,
\end{equation}
where the last relation  follows from the condition that $\psi^{(1)}$ vanishes at the infinity, $x,\,y \to \infty$, where the deformation is zero. Since $\psi^{(1)}$ is a harmonic function, we conclude that the vanishing normal derivative on the  boundary, Eq.~(\ref{eq:25c}), implies that this function vanishes everywhere, that is, $\psi^{(1)}(x,y,z) \equiv 0$ (see e.g.~\cite{Tikhonov}). Hence
\begin{equation}
\label{eq:25d}
\tilde{\sigma}_{ij}^{el(1)}=-E_1z\nabla_i\nabla_j f^{(1)}
\end{equation}
and the third boundary condition in Eq.~(\ref{eq22a}), $\tilde{\sigma}_{zz}^{el(1)}=0$ at $z=0$,   is automatically fulfilled.
Taking into account that the function $f^{(1)}$ is harmonic, we obtain,
\begin{equation}
\label{eq:25e}
\nabla_j \tilde{\sigma}_{ij}^{el(1)}= -E_1  \nabla_i \frac{\partial f^{(1)} }{\partial z}\,.
\end{equation}
Substituting the above relation for $\nabla_j \tilde{\sigma}_{ij}^{el(1)}$ and Eq.~(\ref{eq08}) for $\nabla _j\sigma_{ij}^{v(1)}$ into Eq.~(\ref{eq22}), we recast this equation  into the form,
$$
E_1  \nabla_i \frac{\partial f^{(1)} }{\partial z} =-
\frac{3(E_2 \eta_1-E_1\eta_2)}{(4 E_1+3E_2)} \nabla_i\frac{\partial\dot{f}^{(0)}}{\partial z}\,
$$
which implies the relation between functions $f^{(1)}$ and
$\dot{f}^{(0)}$:
\begin{eqnarray}
\label{eq28}
f^{(1)} \!\!\!\!&=&\!\!\!\!-\alpha_1 \dot{f}^{(0)}\\
\label{eq28al}
\alpha_1\!\!\!\!&=&\!\!\!\!\frac{3(E_2\eta_1-E_1\eta_2)}{E_1(3E_2+4E_1)}
\!=\!\frac{(1\!+\!\nu)(1\!-\!2 \nu)}{(1-\nu)Y} \!\left[\frac{2\!+\!2\nu}{3\!-\!6\nu}\eta_1 - \eta_2\right] \nonumber
\,.
\end{eqnarray}
The function $f^{(1)}$ may be now exploited to express the displacement $\tilde{u}_z^{(1)}$ on the contact plane. Using Eq.~(\ref{eq24}) with $K^{(1)}=-\frac12$ we write for $\tilde{u}_z^{(1)}$:

\begin{equation}
\tilde{u}_z^{(1)} =\frac12 f^{(1)} - \frac{z}{2} \frac{\partial f^{(1)} }{\partial z}\,;
\label{eq28a}
\end{equation}
substituting there $f^{(1)}$ from Eq.~(\ref{eq28}) we arrive at
\begin{equation}
\tilde{u}_z^{(1)} = -\frac12 \alpha_1 \left( \dot{f}^{(0)} -z \frac{\partial \dot{f}^{(0)} }{\partial z} \right)\,,
\label{eq28b}
\end{equation}
where $f^{(0)}$ is given by Eq.~(\ref{eq17}). Thus, the above relation presents the solution for  the displacement $\tilde{u}_z^{(1)}$.  For the contact plane $z=0$ it yields the boundary condition for Eq.~(\ref{eq23}):
\begin{equation}
\left.\bar{u}_z^{(1)}\right|_{z=0} = -\left.\tilde{u}_z^{(1)}\right|_{z=0}= \left.\frac12 \alpha_1 \dot{f}^{(0)}\right|_{z=0}\,.
\label{eq28c}
\end{equation}
Taking into account that $\left. \frac12 \dot{f}^{(0)}\right|_{z=0}=\left.\dot{u}_z^{(0)}\right|_{z=0}$, according to Eq.~(\ref{eq20}), we obtain, expressing $\dot{u}_z^{(0)}$ in terms of $\dot{\sigma}_{zz}^{el(0)}$, as it follows from Eq.~(\ref{eq20a}):

\begin{equation}
\label{eq:29a}
\left.\bar{u}_z^{(1)}\right|_{z=0}=-\frac{(1-\nu^2)}{\pi Y} \int\!\!\! \int_S
\frac{ \alpha_1 \dot{\sigma}_{zz}^{el(0)}(x',y', z=0)\,\mathrm{d}x'\mathrm{d}y'}{ \left|{\bf r} -{\bf r}'\right|}\,.
\end{equation}
Comparing then Eqs.~(\ref{eq:29b}) and (\ref{eq:29a}) we conclude that the first-order stress tensor $\bar{\sigma}_{zz}^{el(1)}$ at the contact plane reads,
\begin{equation}
\label{eq:30a}
\left.\bar{\sigma}_{zz}^{el(1)}\right|_{z=0} = \left. \alpha_1 \dot{\sigma}_{zz}^{el(0)}\right|_{z=0}\, .
\end{equation}
Finally we obtain for the total first-order stress tensor $\sigma_{zz}^{(1)}$:
\begin{eqnarray}
\label{eq:30b}
\left. \sigma_{zz}^{(1)}\right|_{z=0} &=& \left. \left( \bar{\sigma}_{zz}^{el(1)}+ \tilde{\sigma}_{zz}^{el(1)} + \sigma_{zz}^{v(1)} \right) \right|_{z=0} \nonumber \\
&=&\left.(\alpha_0+\alpha_1) \dot{\sigma}_{zz}^{el(0)}\right|_{z=0}\, ,
\end{eqnarray}
where we use Eqs.~(\ref{eq19}) and (\ref{eq:30a}) and take into account that $\tilde{\sigma}_{zz}^{el(1)}=0$ on the contact plane (see Eq.~(\ref{eq22a})).

\section{The dissipative force}
The elastic inter-particles force refers to the zero-order term in
the perturbation expansion (\ref{eq:4}), while the remaining terms
quantify dissipation. Hence, in the linear with respect to the
dissipative constants approximation, the   total dissipative force
reads
$$
 F_z^{v(1)}=\int \!\!\!\int_S\sigma_{zz}^{(1)}(x,y)|_{z=0}\, \mathrm{d}x\, \mathrm{d}y\, ,
$$
so that Eq.~(\ref{eq:30b}) yields,
\begin{eqnarray}
\label{eq:31a}
&&F_z^{v(1)} \!= \!A \frac{\partial}{\partial t} \int \!\!\!\int_S\sigma_{zz}^{el(0)}(x,y)|_{z=0}\, \mathrm{d}x\, \mathrm{d}y  = A \dot{F}_z^{el(0)} \nonumber \\
&& A=\alpha_0 +\alpha_1,
\end{eqnarray}
where $F_z^{el(0)}$ is the normal force corresponding to the elastic reaction of the medium. It is equal to the Hertzian force,  Eq.~(\ref{eq:FH}). Using the expressions (\ref{eq19}) and (\ref{eq28}) for $\alpha_0$ and $\alpha_1$ and Eq.~(\ref{eq:FH}) for the Hertzian force,
we arrive at the final result for the dissipative force:
\begin{eqnarray}
\label{eq:31b}
F_z^{v(1)} &=& \frac 32 A B \sqrt{\xi} \dot{\xi}  \\
A\!\!\!&=&\!\!\!
\frac{1}{Y}\frac{1+\nu }{1-\nu }
 \left[\frac{4}{3}\eta_1(1-\nu +\nu^2 )+\eta_2(1-2\nu )^2\right]. \nonumber
\label{eq:31c}
\end{eqnarray}
Here the constant $B$ depends on the geometry of the colliding bodies and their material properties; for the simple case of a collision of a sphere with a hard plane, it  is given by Eq.~(\ref{eq:FH}). For a collision of two spheres of radii $R_1$ and $R_2$ of the same material it reads~\cite{Landau:1965,bshp96},
$$
B=\frac{2 Y}{3(1-\nu^2)} \sqrt{R_{\rm eff}} \qquad \qquad R_{\rm eff} =\frac{R_1R_2}{R_1+R_2}.
$$
Generally, $B$ depends on the local curvatures of the bodies at the contact, e.g.~\cite{bshp96,Landau:1965,Denisetal2014}. Although the derivation has been illustrated  for the simple case, it remains  valid for the bodies of any convex shapes and different materials~\cite{Denisetal2014}.

Note that the new result (\ref{eq:31b}) for the dissipative force
has been obtained by a rigorous perturbation approach. It contains
all first-order  terms with respect to the small parameter
$\lambda_1$, proportional to the material viscosities $\eta_{1/2}$
which  guarantees the physical consistency of the theory. On the
contrary, the previous result, based on the quasi-static
approximation suffers from the incomplete account of the
first-order stress terms. Indeed, this approximation takes into
account $\hat{\sigma}^{v(1)}$ but ignores $\hat{\sigma}^{el(1)}$.
Physically, $\hat{\sigma}^{v(1)}$ is the component of the stress
associated with the strain rate (i.e. with the relative motion of
different parts of the material) and thus has a ``purely
dissipative'' nature. This stress causes an additional strain in
the bulk, and the respective displacement field ${\bf
u}^{(1)}({\bf r})$ which  gives rise to the excess elastic stress
$\hat{\sigma}^{el(1)}$; both first-order stress terms are of the
same order of magnitude as it follows from Eqs.~(\ref{eq19}),
(\ref{eq:30a}) and (\ref{eq28}). Hence the quasi-static
approximation is not generally valid. It manifests its
inconsistency for the case of $\nu =1/2$, which corresponds to
materials with very small elastic shear module (like rubber).
Although this approximation predicts vanishing dissipation in such
materials, there are no physical mechanisms that could assure the
energy conservation.  At the same time, our new theory is free
from such inconsistencies.

\section{Conclusion}
We develop a mathematically rigorous method to describe the
dissipative force acting between viscoelastic bodies during a
collision. It is based on a perturbation scheme, applied to the
over-damped continuum mechanics equation, with the inertial
effects neglected. We use the small parameter, which is the ratio
of microscopic relaxation time and the characteristic time of a
collision and is proportional to the dissipative constants of the
material. Applying the perturbation approach  we obtain the
dissipative force, linear with respect to this small parameter.
The presented  method is rather general and may be further
developed  to take into account the inertial effects as well as
the high-order corrections with respect to the small parameter.
The obtained dissipation force is expressed in terms of the time
derivative of the elastic force, as it follows from the Hertz
theory, and elastic and viscous material constants. It noticeably
differs from the one obtained previously within the quasi-static
approximation and demonstrates physically correct behavior for the
whole range of material parameters. Finally, we wish to stress
that the proposed approach may be also applied  for similar
continuum mechanics problems, where dissipation in a bulk due to
the strain rate  is addressed.

\begin{acknowledgements}
AVP and DSG acknowledge financial support from the Russian Science
Foundation (grant no.~14-21-00090).
\end{acknowledgements}

\bibliographystyle{eplbib}

\begin{thebibliography}{10}
\expandafter\ifx\csname url\endcsname\relax\def\url#1{\texttt{#1}}\fi

\bibitem{hertz1881}
\Name{Hertz H.} \REVIEW{J. f. Reine u. Angew. Math. }{92}{1881}{156}.

\bibitem{PoeschelSchwager2005}
\Name{Poeschel T. \and Schwager T.} \Book{Computational Granular Dynamics}
  (Springer, Berlin) 2005.

\bibitem{LudingReview2008}
\Name{Luding S.} \REVIEW{Nonlinearity }{22}{2009}{R101}.

\bibitem{Poschl1928}
\Name{Poeschl T.} \REVIEW{Z. Phys. }{46}{1928}{142}.

\bibitem{Poeschel2011}
\Name{Montaine M., Heckel M., Kruelle C., Schwager T. \and Poeschel T.}
  \REVIEW{Phys. Rev. E }{84}{2011}{041306}.

\bibitem{Pao1955}
\Name{Pao Y.-H.} \REVIEW{J. Appl. Phys. }{26}{1955}{1083}.

\bibitem{bshp96}
\Name{Brilliantov N., Spahn F., Hertzsch J. \and P\"oschel T.} \REVIEW{Phys.
  Rev. E }{53}{1996}{5382}.

\bibitem{kk87}
\Name{Kuwabara G. \and Kono K.} \REVIEW{J. Appl. Phys. Part 1
  }{26}{1987}{1230}.

\bibitem{Zheng1}
\Name{Zheng Q.~J., Zhu H.~P. \and Yu A.~B.} \REVIEW{Powder Technology
  }{226}{2012}{130}.

\bibitem{Zheng2}
\Name{Zheng Q.~J., Zhou Z.~Y. \and Yu A.~B.} \REVIEW{Powder Technology
  }{248}{2013}{25}.

\bibitem{Landau:1965}
\Name{Landau L.~D. \and Lifshitz E.~M.} \Book{Theory of Elasticity} (Oxford
  University Press, Oxford) 1965.

\bibitem{bri2}
\Name{Brilliantov N.~V., Albers N., Spahn F. \and P{\"o}schel T.}
  \REVIEW{Phys.\ Rev.\ E }{76}{2007}{051302}.

\bibitem{Dintwa2004}
\Name{Dintwa E., van Zeebroeck M. \and Ramon H.} \REVIEW{Eur. J. Phys. B
  }{39}{2004}{77–85}.

\bibitem{Denisetal2014}
\Name{Goldobin D.~S., Susloparov E.~A., Pimenova A.~V. \and Brilliantov N.}
  \REVIEW{preprint }{}{2014}{}.

\bibitem{book}
\Name{Brilliantov N.~V. \and P\"oschel T.} \Book{Kinetic theory of Granular
  Gases} (Oxford University Press, Oxford) 2004.

\bibitem{Tikhonov}
\Name{Tikhonov A.~N. \and Samarskii A.~A.} \Book{Equations of Mathematical
  Physics} (Dover Publications Inc., New York) 1963.

\end{thebibliography}

\end{document}